 \documentclass{article}

 \def\barr{\left(\begin{array}}
 \def\earr{\end{array}\right)}
 \def\beq#1{\begin{equation}\label{#1}}
 \def\eeq{\end{equation}}
 \def\ber#1{\begin{eqnarray}\label{#1} \nqq}
 \def\eer{\end{eqnarray}}

 \newcommand{\bear}[1]{\begin{eqnarray}\label{#1}}
 \newcommand{\ear}{\end{eqnarray}}

 \renewcommand{\theequation}{\arabic{section}.\arabic{equation}}
 \catcode`\@=11 \@addtoreset{equation}{section}\catcode`\@=12

 \newcommand{\N}{ {\bf N} }
 \newcommand{\R}{ {\bf R} }

 \newcommand{\sign}{\mathop{\rm sign}\nolimits}
 
 \newcommand{\btu}{\bigtriangleup}
 \newcommand{\eps}{\varepsilon}
 \newcommand{\tri}{\triangle}
 
 \newcommand{\p}{\partial}
 \newcommand{\nn}{\nonumber}
 
 \newcommand{\fnm}{\footnotemark}
 \newcommand{\fnt}{\footnotetext}

\begin{document}

 \begin{center} \large \bf

  S-BRANE SOLUTIONS WITH ACCELERATION\\
  IN MODELS WITH FORMS AND MULTIPLE EXPONENTIAL POTENTIALS

 \end{center}

 \vspace{1.03truecm}

 \bigskip

 \begin{center}

  \normalsize\bf
  V. D. Ivashchuk\fnm[1]\fnt[1]{ivas@rgs.phys.msu.su},
  V. N. Melnikov\fnm[2]\fnt[2]{melnikov@rgs.phys.msu.su}

  \bigskip
  
  \it
  Center for Gravitation and Fundamental Metrology,
  VNIIMS, 3-1 M. Ulyanovoy Str., Moscow, 119313, Russia and

  Institute of Gravitation and Cosmology,
 Peoples' Friendship University of Russia,
 6 Miklukho-Maklaya St., Moscow 117198, Russia

  \bigskip
  
  \normalsize\bf  
  and S.-W. Kim\fnm[3]\fnt[3]{sungwon@mm.ewha.ac.kr}

 \bigskip
  
  \it
  Department of Science Education and Basic Science Research Institute,
  Ewha Woman's University, Seoul 120-750, Korea

 \end{center}

 \vspace{15pt}

 \begin{abstract}

 A family of generalized $S$-brane solutions
 with orthogonal intersection rules
 and  $n$ Ricci-flat factor spaces in the theory with several scalar
 fields, antisymmetric forms and multiple scalar potential
 is  considered.  Two subclasses of solutions with power-law and
 exponential behaviour of scale factors are singled out.
 These subclasses contain sub-families of solutions with
 accelerated expansion of certain factor spaces.
 Some examples of solutions with  exponential dependence of one scale
 factor and constant scale factors of "internal" spaces (e.g.
 Freund-Rubin type solutions) are also considered.

 \end{abstract}


 \pagebreak

\section{Introduction}

Recently, the discovery of the cosmic acceleration
 \cite{Riess,Perl} was a starting point
for a big number of publications on
multidimensional cosmology giving some explanations of this phenomenon
using certain multidimensional models \cite{Me}, e.g. those of superstring or
supergravity origin (see, for example, \cite{IMSjhep,Neu,Leb} and
references therein).
These solutions deal with time-dependent scale factors of internal spaces
(for reviews see \cite{IMJ,Sam,IMtop,Ierice}) and contain as a special case
the so-called S-brane solutions \cite{S1}, i.e. space like
analogues of $D$-branes \cite{Polc}, see for example
 \cite{S2,S3,S4,S5,Isbr,Ohta,Iohta} and references therein.
For earlier $S$-brane solutions see also \cite{BF,LOW,LMPX}.

In our recent paper \cite{IMSjhep} we have obtained
a family of cosmological solutions
with  $(n+1)$ Ricci-flat spaces in the theory with several scalar
fields and multiple exponential potential when
coupling vectors in exponents obey certain
 "orthogonality" relations. In \cite{IMSjhep}
two subclasses of "inflationary-type" solutions with power-law
and exponential behaviour of scale factors were found
and solutions with accelerated expansion  were singled out.

 2-component models in many dimensions having the
acceleration were found also for different matter sources in our earlier
papers:  with the cosmological constant in \cite{a2}, with a perfect fluid
in \cite{a1}, with 2 non-Ricci-flat spaces \cite{a3}, with p-branes and
static internal spaces in \cite{a4} and in four dimensions with a perfect
fluid and a scalar field with the exponential potential in \cite{a6,a7},
using methods, developed in our multidimensional approach. In \cite{GMX} we
showed that the cosmic acceleration and coincidence problems may be solved
by using an x-fluid as a quintessence and a viscous fluid as a normal
matter. Viscosity of the normal matter in this case can be explained by its
own multicomponent structure.

In this paper we generalize "inflationary-type"
solutions  from  \cite{IMSjhep} to
 $S$-brane configurations in the model with
antisymmetric forms and scalar fields. Two subclasses of these
solutions with the power-law and  exponential  behaviour of scale factors in
the synchronous time are singled out. These subclasses contain sub-families
of solutions with accelerated expansion of certain factor spaces.

\newpage

Here we deal with a model governed by the action
  \bear{1.1}
   S_g= \int d^Dx
   \sqrt{|g|}\biggl\{R[g]-h_{\alpha\beta}g^{MN}\p_M\varphi^\alpha
   \p_N\varphi^\beta
  \\ \nonumber
  - \sum_{a\in\tri}\frac{\theta_a}{n_a!}
   \exp[2\lambda_a(\varphi)](F^a)^2
   - 2 V_{\varphi}(\varphi) \biggr\}
  \ear
where $g=g_{MN}(x)dx^M\otimes dx^N$ is a metric,
 $\varphi=(\varphi^\alpha)\in\R^l$ is a vector of scalar fields,
 $(h_{\alpha\beta})$ is a  constant symmetric
non-degenerate $l\times l$ matrix $(l\in \N)$,
 $\theta_a=\pm1$,
 \beq{1.2a}
 F^a =    dA^a
 =  \frac{1}{n_a!} F^a_{M_1 \ldots M_{n_a}}
 dz^{M_1} \wedge \ldots \wedge dz^{M_{n_a}}
 \eeq
is a $n_a$-form ($n_a\ge1$), $\lambda_a$ is a
1-form on $\R^l$: $\lambda_a(\varphi)=\lambda_{\alpha a}\varphi^\alpha$,
 $a\in\tri$, $\alpha=1,\dots,l$.
In (\ref{1.1})
we denote $|g| =   |\det (g_{MN})|$,
 \beq{1.3a}
  (F^a)^2_g  =
  F^a_{M_1 \ldots M_{n_a}} F^a_{N_1 \ldots N_{n_a}}
  g^{M_1 N_1} \ldots g^{M_{n_a} N_{n_a}},
 \eeq
 $a \in \tri$.
Here $\tri$ is some finite set.
For pseudo-Euclidean metric of signature $(-,+, \ldots,+)$
all $\theta_a = 1$.
Here
  \beq{pot}
    V_{\varphi}(\varphi) =
    \sum_{s \in S_{pot}} \Lambda_s \exp[ 2 \lambda_s (\varphi) ]
  \eeq
is the scalar potential  ($S_{pot}$ is non-empty finite set).
The case $V_{\varphi}(\varphi) = 0$
was considered recently in \cite{IMSgrg}.

The paper is organized as following.
In Section 2 we consider cosmological-type solutions with composite
intersecting $S$-branes for the model with scalar fields and fields of forms
in  the  presence of multiple scalar potential. These
solutions generalize that from \cite{IMtop,Ierice,Isbr,Iohta}
obtained for  zero scalar potential.  Section 3 is
devoted to exceptional ("inflationary-type") $S$-brane solutions.
In Section 4 we consider a class of static solutions
defined on product of $n$ Einstein spaces
 $N_i$ generalizing that of \cite{I}.  For certain first factor-space
 $N_1 = \R \times M_1$ of non-zero curvature a solution from this class may
be considered as a cosmological solution with exponential dependence of
scale factor for Ricci-flat submanifold  $M_1$ and static
internal spaces $N_i$, $i > 1$.


 \section{Cosmological-type solutions with composite
 intersecting $p$-branes}

 \subsection{Solutions with $n$ Ricci-flat spaces}

Here we  consider a family of
solutions to field equations corresponding to the action
 (\ref{1.1}) and depending upon one variable $u$
and generalizing that from \cite{Ierice,Isbr,Iohta} defined for
 $V_{\varphi} = 0$ (see also \cite{IK,IMtop}).

These solutions are defined on the manifold
  \beq{1.2}
  M =    (u_{-}, u_{+})  \times
  M_1  \times M_2 \times  \ldots \times M_{n},
  \eeq
where $(u_{-}, u_{+})$  is  an interval belonging to $\R$,
and have the following form
 \bear{1.3}
  g= \biggl(\prod_{s \in S} [f_s(u)]^{2 d(I_s) h_s/(D-2)} \biggr)
  \biggr\{ \exp(2c^0 u + 2 \bar c^0) w du \otimes du  + \\ \nn
  \sum_{i = 1}^{n} \Bigl(\prod_{s\in S}
  [f_s(u)]^{- 2 h_s  \delta_{i I_s} } \Bigr)
  \exp(2c^i u+ 2 \bar c^i) \hat{g}^i \biggr\}, \\ \label{1.4}
  \exp(\varphi^\alpha) =
  \left( \prod_{s\in S} f_s^{h_s \chi_s \lambda_{a_s}^\alpha} \right)
  \exp(c^\alpha u + \bar c^\alpha), \\ \label{1.5}
  F^a= \sum_{s \in S} \delta^a_{a_s} {\cal F}^{s},
 \ear
 $\alpha=1,\dots,l$; $a \in \tri$.

In  (\ref{1.3})  $w = \pm 1$,
 $g^i=g_{m_i n_i}^i(y_i) dy_i^{m_i}\otimes dy_i^{n_i}$
is a Ricci-flat  metric on $M_{i}$, $i=  1,\ldots,n$,
  \beq{1.11}
   \delta_{iI}=  \sum_{j\in I} \delta_{ij}
  \eeq
is the indicator of $i$ belonging
to $I$: $\delta_{iI}=  1$ for $i\in I$ and $\delta_{iI}=  0$ otherwise.

Here
  \beq{1.set}
   S \equiv S_{br} \sqcup S_{pot}
     \eeq
is a union of two non-intersecting sets $S_{br}$ and $S_{pot}$,
describing branes and potential terms, respectively.
(Here and in what follows $\sqcup$ means the union
of non-intersecting sets.) The  $p$-brane  set  $S$ is by definition
  \beq{1.6}
  S_{br} =  S_e \sqcup S_m, \quad
  S_v=  \sqcup_{a\in\tri}\{a\}\times\{v\}\times\Omega_{a,v},
  \eeq
 $v=  e,m$ and $\Omega_{a,e}, \Omega_{a,m} \subset \Omega$,
where $\Omega =   \Omega(n)$  is the set of all non-empty
subsets of $\{ 1, \ldots,n \}$.
Any $p$-brane index $s \in S_{br}$ has the form
  \beq{1.7}
   s =  (a_s,v_s, I_s),
  \eeq
where
 $a_s \in \tri$ is colour index, $v_s =  e,m$ is electro-magnetic
index and the set $I_s \in \Omega_{a_s,v_s}$ describes
the location of $p$-brane worldvolume.

The sets $S_e$ and $S_m$ define electric and magnetic
 $p$-branes, correspondingly. In (\ref{1.4})
  \beq{1.8}
   \chi_s  =  +1,
  \eeq
for $s \in S_e$ and
   \beq{1.8pot}
     \chi_s  =  -1,
   \eeq
for $s \in S_m \cup S_{pot}$.

In (\ref{1.5})  forms
  \beq{1.9}
  {\cal F}^s= Q_s  f_{s}^{- 2} du \wedge\tau(I_s),
  \eeq
 $s\in S_e$, correspond to electric $p$-branes and
forms
  \beq{1.10}
  {\cal F}^s= Q_s \tau(\bar I_s),
  \eeq
  $s \in S_m$,
correspond to magnetic $p$-branes; $Q_s \neq 0$, $s \in S_{br}$.
Here  and in what follows
  \beq{1.13a}
  \bar I \equiv I_0 \setminus I, \qquad I_0 = \{1,\ldots,n \}.
  \eeq

All manifolds $M_{i}$ are assumed to be oriented and
connected and  the volume $d_i$-forms
  \beq{1.12}
  \tau_i  \equiv \sqrt{|g^i(y_i)|}
  \ dy_i^{1} \wedge \ldots \wedge dy_i^{d_i},
  \eeq
and parameters
  \beq{1.12a}
   \varepsilon(i)  \equiv {\rm sign}( \det (g^i_{m_i n_i})) = \pm 1
  \eeq
are well-defined for all $i=  1,\ldots,n$.
Here $d_{i} =   {\rm dim} M_{i}$, $i =   1, \ldots, n$;
 $D =   1 + \sum_{i =   1}^{n} d_{i}$. For any
set $I =   \{ i_1, \ldots, i_k \} \in \Omega$, $i_1 < \ldots < i_k$,
we denote
  \bear{1.13}
  \tau(I) \equiv \hat{\tau}_{i_1}  \wedge \ldots \wedge
  \hat{\tau}_{i_k},
  \\
  \label{1.15}
  d(I) \equiv   \sum_{i \in I} d_i, \\
  \label{1.15a}
  \varepsilon(I) \equiv \varepsilon(i_1) \ldots \varepsilon(i_k).
 \ear


The parameters  $h_s$ appearing in the solution
satisfy the relations
 \beq{1.16}
  h_s = (B_{s s})^{-1},
 \eeq
where
 \beq{1.17}
  B_{ss'} \equiv
   d(I_s\cap I_{s'})+\frac{d(I_s)d(I_{s'})}{2-D}+
  \chi_s \chi_{s'} \lambda_{\alpha s}\lambda_{\beta s'}
  h^{\alpha\beta},
 \eeq
 $s, s' \in S$, with $(h^{\alpha\beta})=(h_{\alpha\beta})^{-1}$.
Here  $I_s = I_0 = \{1,\ldots,n \} $ for $s \in S_{pot}$
and $\lambda_{\alpha s} = \lambda_{\alpha a_s}$ for $s \in S_{br}$.

We assume that
 \beq{1.17a}
 ({\bf i}) \qquad B_{ss} \neq 0,
 \eeq
for all $s \in S$, and
 \beq{1.18b}
 ({\bf ii}) \qquad B_{s s'} = 0,
 \eeq
for $s \neq s'$, i.e. canonical (orthogonal) intersection rules
are satisfied (see \cite{IMC,IMJ} and subsection 2.2 below).

The moduli functions read
 \bear{1.4.5}
  f_s(u)=
  R_s \sinh(\sqrt{C_s}(u-u_s)), \;
  C_s>0, \; h_s \eps_s<0; \\ \label{1.4.7}
  R_s \sin(\sqrt{|C_s|}(u-u_s)), \;
  C_s<0, \; h_s\eps_s<0; \\ \label{1.4.8}
  R_s \cosh(\sqrt{C_s}(u-u_s)), \;
  C_s>0, \; h_s\eps_s>0; \\ \label{1.4.9}
  P_s |h_s|^{-1/2}(u-u_s), \; C_s=0, \; h_s\eps_s<0,
  \ear
where  $C_s$, $u_s$  are constants, $s \in S$.
 Here we denote
 \bear{R.1}
 R_s = |Q_s|| h_s C_s|^{-1/2}, \quad s \in S_{br},
  \\  \label{R.2}
 R_s = \sqrt{2 |\Lambda_s|} |h_s C_s|^{-1/2},
 \quad s \in S_{pot},
 \ear
 $P_s = |Q_s|$ for $s \in S_{br}$ and
 $P_s = \sqrt{2 |\Lambda_s|}$ for $s \in S_{pot}$.

 We  also denote
 \beq{1.22a}
    \eps_s = - \sign(w \Lambda_s)
 \eeq
for $s \in S_{pot}$  and (as in  \cite{IMtop})
  \beq{1.22}
   \eps_s=(-\eps[g])^{(1-\chi_s)/2}\eps(I_s) \theta_{a_s},
  \eeq
for  $s \in S_{br}$, where  $\eps[g]\equiv\sign(\det(g_{MN}))$.
 More explicitly
 (\ref{1.22}) reads: $\eps_s=\eps(I_s) \theta_{a_s}$ for
 $v_s = e$ and $\eps_s=-\eps[g] \eps(I_s) \theta_{a_s}$  for
 $v_s = m$.

Vectors $c=(c^A)= (c^i, c^\alpha)$ and
 $\bar c=(\bar c^A)$ obey the following constraints \cite{IMJ}
 \beq{1.27}
  \sum_{i \in I_s}d_ic^i-\chi_s\lambda_{s \alpha}c^\alpha=0,
  \qquad
  \sum_{i\in I_s}d_i\bar c^i-
  \chi_s\lambda_{s \alpha}\bar c^\alpha=0, \quad s \in S,
   \eeq
  \bear{1.30aa}
  c^0 = \sum_{j=1}^n d_j c^j,
  \qquad
  \bar  c^0 = \sum_{j=1}^n d_j \bar c^j,
  \\  \label{1.30a}
  \sum_{s \in S} C_s  h_s +
    h_{\alpha\beta}c^\alpha c^\beta+ \sum_{i=1}^n d_i(c^i)^2
  - \left(\sum_{i=1}^nd_ic^i\right)^2 = 0.
 \ear

Here we identify notations
 $\hat{g}^{i} = p_{i}^{*} g^{i}$ is the
pullback of the metric $g^{i}$  to the manifold  $M$ by the
canonical projection: $p_{i} : M \rightarrow  M_{i}$, $i = 1,
 \ldots, n$. An analogous
notations are kept for volume forms etc.

Due to (\ref{1.9}) and  (\ref{1.10}), the dimension of
 $p$-brane worldvolume $d(I_s)$ is defined by
 \beq{1.16a}
  d(I_s)=  n_{a_s}-1, \quad d(I_s)=   D- n_{a_s} -1,
 \eeq
for $s \in S_e, S_m$, respectively.
For $s \in S_{pot}$  we get domain walls with $d(I_s)=  D - 1$ (see
Appendix).  (The $p$-brane parameter reads: $p =   p_s =   d(I_s)-1$).

 {\bf Restrictions on $p$-brane configurations.}
The solutions  presented above are valid if two
restrictions on the sets of composite $p$-branes are satisfied \cite{IK}.
These restrictions
guarantee  the block-diagonal form of the  energy-momentum tensor
and the existence of the sigma-model representation (without additional
constraints) \cite{IMC}.

The first restriction reads
 \beq{1.3.1a}
  {\bf (R1)} \quad d(I \cap J) \leq d(I)  - 2,
 \eeq
for any $I,J \in\Omega_{a,v}$, $a\in\tri$, $v= e,m$
(here $d(I) = d(J)$).

The second restriction is following one
 \beq{1.3.1b}
  {\bf (R2)} \quad d(I \cap J) \neq 0,
 \eeq
for $I\in\Omega_{a,e}$ and $J\in\Omega_{a,m}$, $a \in \tri$.

The  derivation of the solution is presented in  Appendix B.

 {\bf $S$-branes.} The space-like brane  solutions (or $S$-branes)
 appear when the  metric  ({\ref{1.3}}) has a  signature $(-,+,
 \ldots,+)$,  $w = -1$ and all metrics $g^i$ have Euclidean
 signatures. In this case all $\theta_a = 1$
 and
  \beq{1.22c}
   \eps_s=  1,
  \eeq
 for all $s \in S_{br}$.

 \subsection{Minisuperspace covariant relations}

 Here we present the minisuperspace covariant relations
 from \cite{IMJ,IMC} for completeness. Let
  \beq{2.1}
  (\bar{G}_{AB})=\barr{cc}
  G_{ij}& 0\\
  0& h_{\alpha\beta}
  \earr,
 \qquad
 (\bar G^{AB})=\left(\begin{array}{cc}
 G^{ij}&0\\
 0&h^{\alpha\beta}
 \end{array}\right)
  \eeq
be, correspondingly,
a (truncated) target space metric and inverse to it,
where  (see \cite{IMZ})
 \beq{2.2}
   G_{ij}= d_i \delta_{ij} - d_i d_j, \qquad
   G^{ij}=\frac{\delta^{ij}}{d_i}+\frac1{2-D},
 \eeq
and
 \beq{2.3}
   U_A^s c^A =
   \sum_{i \in I_s} d_i c^i - \chi_s \lambda_{s \alpha} c^{\alpha},
   \quad
   (U_A^s) =  (d_i \delta_{iI_s}, -\chi_s \lambda_{s \alpha}),
 \eeq
are co-vectors, $s \in S$ and $(c^A)= (c^i, c^\alpha)$.

The scalar product from \cite{IMC} reads
 \beq{2.4}
  (U,U')=\bar G^{AB} U_A U'_B,
 \eeq
for $U = (U_A), U' = (U'_A) \in \R^N$, $N = n + l$.

The scalar products  for vectors
 $U^s$  were calculated in \cite{IMC}
 \beq{2.7}
  (U^s,U^{s'})= B_{s s'},
 \eeq
where  $s, s'$ belong to $S$ and
 $B_{s s'}$ are defined in (\ref{1.17}).
Due to relations (\ref{1.18b}) $U^s$-vectors
are orthogonal, i.e.
 \beq{2.7a}
 (U^s,U^{s'})= 0,
 \eeq
for $s \neq s'$.

The linear and quadratic constraints
from (\ref{1.27}) and (\ref{1.30a}),
respectively, read in minisuperspace covariant
form as follows:
 \beq{2.8}
   U_A^s c^A = 0, \qquad U_A^s \bar{c}^A = 0,
   \eeq
  $s \in S$,
and
 \beq{2.10}
  \sum_{s \in S} C_s  h_s +
  \bar G_{AB} c^A c^B = 0.
 \eeq

  \section{Special solutions}

Now we consider a special case of classical solutions
from the previous section when $C_s = u_s = c^i = c^{\alpha} = 0$
and
\beq{6.0}
   B_{ss} \eps_s < 0,
\eeq
$s \in S$.

We get two families of solutions written in synchronous-type
variable with:

  A) power-law dependence of scale factors for $B \neq -1$,

  B) exponential dependence of scale factors for $B  = -1$,

where

 \beq{6.1}
  B = \sum_{s \in S} h_s \frac{d(I_s)}{D - 2}.
 \eeq

Remind that

 \beq{6.1a}
  h_s^{-1} =   B_{ss} =
   d(I_s)\left(1 -  \frac{d(I_s)}{D -2}  \right) +
  \lambda_{\alpha s}\lambda_{\beta s}   h^{\alpha\beta}.
 \eeq

\subsection{Power-law solutions}

Let us consider the solution corresponding to
the case $B   \neq -1$. The solution reads
\bear{6.2}
  g=  w d\tau \otimes d\tau
  +   \sum_{i = 1}^{n} A_i \tau^{2 \nu_i} \hat{g}^i, \\
  \label{6.3}
  \varphi^\alpha= \frac{1}{B+1}
  \sum_{s \in S} \chi_s h_s \lambda_{s}^{\alpha} \ln \tau
  + \varphi^\alpha_0,
\ear
where $\tau > 0$,
\beq{6.4}
  \nu_{i} = - \frac{1}{B+1} \sum_{s \in S} h_s
  \left(\delta_{iI_s} - \frac{d(I_s)}{D-2} \right),
\eeq
$i = 1, \dots, n$.

Here
\beq{6.5}
|h_s| \left( \prod_{i \in \bar{I}_s} A_i^{d_i} \right)
  \exp( 2 \chi_s \lambda_{s \alpha} \varphi^{\alpha}_0)
  = Q_s^2 |B+1|^{2},
\eeq
$s \in S_{br}$ and
\beq{6.5a}
|h_s| \exp( 2 \chi_s \lambda_{s \alpha} \varphi^{\alpha}_0)
  = 2 |\Lambda_s| |B+1|^{2},
\eeq
$s \in S_{pot}$.  $A_i > 0$ are arbitrary constants.

The elementary forms read
  \beq{6.9}
  {\cal F}^s=  \frac{|h_{s}| A^{1/2}}{Q_s (B+1)|B+1|}
\tau^{-(B+2)/(B+1)}  d\tau \wedge \tau(I_s),
  \eeq
$s \in S_e $ (for electric case) and
forms
  \beq{6.10}
  {\cal F}^s= Q_s \tau(\bar I_s),
  \eeq
  $s \in S_m$ (for magnetic case). Here
  and in what follows  $w = \pm 1$, $Q_s \neq 0$, $s \in S_{br}$,
  $A = \prod_{i=1}^{n} A_i^{d_i}$ and $h_s$ is defined in
  (\ref{6.1a}).

\subsubsection{Solutions with accelerated expansion of $M_1$}

In what follows we are interested in cosmological solutions ($w  = -1$) with
accelerated expansion of the first factor space $M_1$ that, evidently,
takes place if and only if

 \beq{6.11}
     \nu_1 > 1.
 \eeq

 Let us consider the case of one brane: $S = \{ s \}$.

A) First, we put
 \beq{6.12}
     1 \notin I_s,
 \eeq
 i.e. $M_1$ factor space does not belong to the brane.
In this case $d(I_s) \neq D - 1$.

 We get
  \beq{6.13}
     \nu_1 = \frac{1}{1 + (D-2) (h_s d(I_s))^{-1} }.
 \eeq

Due to (\ref{6.11}) the acceleration takes place only if $h_s < 0$, or
(see (\ref{6.1a}))
 \beq{6.14}
   \lambda_{s}^2 <  - d(I_s)\left(1 -  \frac{d(I_s)}{D -2}  \right) < 0,
 \eeq
where here and in what follows
 $\lambda_{s}^2 = \lambda_{\alpha s} \lambda_{\beta s}  h^{\alpha  \beta}$.

From the signature restriction (\ref{6.0}) and $h_s < 0$ we get
  $\eps_s > 0$.

B) Now, we consider another case
 \beq{6.12b}
     1 \in I_s,
 \eeq
 i.e. $M_1$ factor-space belongs to the brane.

We get
  \beq{6.13b}
     \nu_1 = \frac{\Delta_s - 1}{h_s^{-1} + \Delta_s },
    \qquad
    \Delta_s = d(I_s)/(D - 2).
  \eeq

Due to (\ref{6.11}) the acceleration takes place only if
 \beq{6.14b}
   \frac{h_s^{-1} + 1}{h_s^{-1} + \Delta_s } < 0.
 \eeq
This implies $\Delta_s \neq 1$, or  $d(I_s) \neq D - 2$.

There are two subcases: B1) $d(I_s) < D - 2$ and B2) $d(I_s) = D - 1$.

For B1) $d(I_s) < D - 2$ we get
  \beq{6.15b}
      - 1  < h_s^{-1} < -  \Delta_s,
  \eeq
or, equivalently,
  \beq{6.14be}
   - d(I_s)\left(1 -  \frac{d(I_s)}{D -2}  \right) - 1
     < \lambda_{s}^2 <
   - d(I_s)\left(1 -  \frac{(d(I_s) - 1)}{D -2}  \right) < 0.
   \eeq

 For B2) $d(I_s) = D - 1$  we find
  \beq{6.15c}
     - \Delta_s  < h_s^{-1} < - 1,
  \eeq
 or, equivalently,
  \beq{6.14c}
   0  < \lambda_{s}^2 < \frac{1}{D - 2}.
  \eeq

In both subcases  $h_s < 0$ and hence $ \eps_s > 0$ (due to (\ref{6.0})).
(For $s \in S_{pot}$ this means that $\Lambda_s > 0$.)

The negative value of $\lambda_s^2$ means that the matrix
$(h_{\alpha  \beta})$ is not  positive definite. Thus,
for positive definite $(h_{\alpha  \beta})$,
the one brane solution may describe
the accelerated expansion of certain (say, $M_1$) factor
space only in the  domain wall
case   $d(I_s) = D - 1$ with  coupling constant obeying (\ref{6.14c}).

{\bf Remark.} In the case  of several electric $S$-branes of maximal
dimension $d(I_s) = D - 1$ (i.e. domain walls) the metric
and scalar fields are coinciding (up to notations)
with the solutions obtained in \cite{IMSjhep}
when signature restrictions (\ref{6.0}) are obeyed.

\subsection{Solutions with exponential scale factors}

Here we consider the solution corresponding to
the case $B = -1$. The solution reads
  \bear{6.15d}
    g=  w d\tau \otimes d\tau
    +  \sum_{i = 1}^{n} A_i \exp(2 M \mu_i \tau) \hat{g}^i, \\
       \label{6.16}
    \varphi^\alpha= - M \tau \sum_{s \in S}
    h_s \chi_s \lambda_{s}^{\alpha}
    + \varphi^\alpha_0,
  \ear
where
  \beq{6.17}
    Q_s^2 \exp(- 2 \chi_s \lambda_{s \alpha} \varphi^{\alpha}_0)
    = |h_s| M^2 \prod_{i \in \bar{I}_s} A_i^{d_i},
  \eeq
  $s \in S_{br}$ and
  \beq{6.17a}
    2 |\Lambda_s| \exp(- 2 \chi_s \lambda_{s \alpha} \varphi^{\alpha}_0)
    = |h_s| M^2,
  \eeq
  $s \in S_{pot}$.
  Here
  \beq{6.18}
  \mu_{i} = \sum_{s \in S} h_s
  \left(\delta_{iI_s} - \frac{d(I_s)}{D-2} \right),
  \eeq
 $M$ is parameter
 and $A_i > 0$ are arbitrary constants, $i = 1, \dots, n$.

The elementary forms read
  \beq{6.19}
  {\cal F}^s=  \frac{|h_{s}| A^{1/2}}{Q_s}
   M^2 e^{M \tau} d\tau \wedge \tau(I_s),
  \eeq
for $s \in S_e$,  and
  ${\cal F}^s= Q_s \tau(\bar I_s)$ for $s \in S_m$.

In the cosmological case $w = -1$ we get an accelerated
expansion of the factor space $M_1$ if and only if $\mu_1 M > 0$.
 For any $\mu_1 \neq 0$ this may be achieved by a suitable choice of
 the sign of the parameter $M$.

  \subsubsection{Example: magnetic $S$-brane and domain-wall.}

 Let us consider $D$-dimensional model
 with one scalar field, one antisymmetric form and one exponential potential
 term described by the action
  \beq{e.1}
   S_g= \int d^Dx
   \sqrt{|g|} \biggl\{ R[g]-g^{MN} \p_M \varphi \p_N \varphi
   - \frac{1}{n_1!} \exp(2 \lambda_1 \varphi) F^2
    - 2 \Lambda \exp(2 \lambda_2 \varphi)   \biggr\}
  \eeq
defined on the manifold
  \beq{e.2}
  M =    \R  \times   M_1  \times M_2.
  \eeq
Let the rank of the form $F = dA$ be coinciding with the
dimension of the second factor-space, i.e. $n_1 = d_2$ and
  \beq{e.3}
  \lambda_1 = \pm \frac{d_1}{\sqrt{D-2}}, \qquad
  \lambda_2 = \pm \frac{1}{\sqrt{D-2}},
  \eeq
$D = d_1 + d_2 + 1$
and $(M_1, g_1)$ be of Euclidean
signature $(+, \ldots, +)$
and $(M_2, g_2)$ be of pseudo-Euclidean one
$(-,+, \ldots, +)$. Dilatonic couplings satisfy
the orthogonality relation (\ref{1.18b}),
$B = -1$ and signature restrictions (\ref{6.0}).

The solution with magnetic $S$-brane (corresponding to
$M_1$ factor-space) and domain-wall reads:

\bear{e.15}
    g=  - d\tau \otimes d\tau
    +  \sum_{i = 1}^{2} A_i \exp(2 M d_1^{-1} \delta_i^1 \tau) \hat{g}^i, \\
    \label{e.16}
    \varphi = {\rm const},
\ear
where
\beq{e.17}
 2 \Lambda \exp(2 \lambda_2 \varphi) = M^2,
 \qquad  Q_1^2 \exp(2 \lambda_1 \varphi) =
 \frac{1}{d_1} M^2 A_2^{d_2},
\eeq
$M$ is parameter
and $A_i > 0$ are arbitrary constants, $i = 1, 2$.
The form reads
  \beq{e.19}
    F= Q_1 \hat{\tau}_2.
  \eeq
Remind that $\tau_2 = dvol[g_2]$ is the volume form of
the $(M_2, g_2)$ space.

The "internal" factor space $(M_2, g_2)$ has a constant
scale factor. This subspace contains an extra time direction.

We note that in our case   $B_{s_1 s_1} = d_1$,
   $B_{s_2 s_2} = -1$ and $\eps_{s_1} < 0$,
   $\eps_{s_2} > 0$, where index $s_1$ corresponds
   to the magnetic brane and $s_2$ to the domain wall.

In the next section we shall show that  this solution,
describing  an accelerated expansion of
``our space'' $M_1$ with  static ``internal'' space $M_2$, is a special case
of a more general class of solutions.

\section{Freund-Rubin type solutions}

Here we present a class of static solutions defined
on a product of several manifolds
 \beq{s.1}
     M = N_{1} \times \ldots \times N_{k}
 \eeq
and generalizing that of ref. \cite{a4,I}.

The metric of the solution has the following form
 \beq{s.2}
    g= \hat{h}_1 + \ldots + \hat{h}_k,
 \eeq
where  $h_i$  is an Einstein metric on $N_{i}$  satisfying the relation
 \beq{s.3}
    {\rm Ric}[h_i]= \xi_{i} h_i,
 \eeq
   $\xi_{i} = {\rm const} $, $i=1,\ldots,k$.

Here ${\rm Ric}[h_i]$ is Ricci-tensor corresponding
to $h_{i}$ and $\hat{h}_{i} = \pi_{i}^{*} h_{i}$ is the
pullback of the metric $h_i$  to the manifold  $M$ by the
canonical projection: $\pi_{i} : M \rightarrow  N_{i}$, $i = 1,
 \ldots, k$. Thus, all $(N_{i}, h_{i})$  are Einstein spaces.

The fields of forms and  scalar fields are the following
\bear{s.9}
  F^a = \sum_{I \in \Omega_{a}} Q_{aI} \tau(I),
  \\ \label{s.10}
 \varphi^{\alpha} = {\rm const},
\ear
where $Q_{aI}$ are constants,
$\Omega_{a} \subset \{1, \dots, k \}$ are  subsets,
satisfying the relations
 \beq{s.23}
  d(I) = n_a ,
 \eeq
  $I \in \Omega_a$, $a \in \Delta$, and
the {\bf Restriction } presented below.
The  parameters of the solution obey the relations
 \bear{s.11}
   \sum_{a \in \Delta} \theta_a \lambda^{\alpha}_a
   e^{2 \lambda_{a}(\varphi)}
   \sum_{I \in \Omega_a} Q_{aI}^2 \eps(I)
  + 2 \sum_{s \in S_{pot}}
 \lambda^{\alpha}_s   \Lambda_s  e^{2 \lambda_s (\varphi)}
  = 0, \quad \\
 \label{2.12}
  \xi_i =   \frac{2 }{D-2}
  \sum_{s \in S_{pot}}  \Lambda_s  e^{2 \lambda_s (\varphi)}
+  \sum_{a \in \Delta} \theta_a e^{2 \lambda_{a}(\varphi)}
 \sum_{I \in \Omega_a} Q_{aI}^{2} \eps(I) \left[ \delta^i_I
 - \frac{n_a -1}{D - 2} \right],
 \ear
 $i = 1, \ldots, k$.

The solution is valid if the following restriction
on the sets $\Omega_{a}$, $a \in \Delta$, is satisfied \cite{I}.

{\bf Restriction.}
For any  $a \in \Delta$ and $I,J \in \Omega_{a}$, $I \neq J$, we put
\beq{2.r}
    d(I \cap  J) \leq n_a - 2.
\eeq

These solutions  just follow from the equations of motion
presented in  Appendix A.

The solution (\ref{e.15})-(\ref{e.19})
of the previous section
may be considered as a special solution
defined on the product of two Einstein spaces
$N_1 = \R  \times   M_1$ and $N_2= M_2$
with metrics $h_1 =  - d\tau \otimes d\tau
+  \exp(2 d_1^{-1} M \tau) \hat{g}^1$ ($d_1 = {\rm dim }M_1$) and
$h_2 = g_2$, respectively. Here
$\xi_1 = d_1^{-1} M^2$ and $\xi_2 = 0$.

It should be noted that any  solution of this section
containing an Einstein space of positive curvature
$N_1 = \R  \times   M_1$,
 may be interpreted as a cosmological
solution with static internal spaces and exponentially
expanding Ricci-flat $(M_1, g_1)$ submanifold, if
$h_1 =  - d\tau \otimes d\tau
+  \exp(2 H \tau) \hat{g}^1$ and $\xi_1 = d_1 H^2$
(here $d_1 = {\rm dim }M_1$). In subsection
3.2.1  $Q_{1I}^2 = Q_1^2 A_2^{- d_2} $ and
$I = \{2 \}$.

\section{Conclusions}

In this paper we obtained  generalized $S$-brane solutions
with orthogonal intersection rules
and  $n$ Ricci-flat factor spaces in the theory with several scalar
fields, antisymmetric forms and multiple exponential potential.
We singled out two  subclasses of solutions
with power-law and  exponential behaviour of scale factors
depending in general on charge densities of branes, their dimensions,
intersections and dilatonic couplings.

These subclasses contain sub-families of solutions with accelerated
expansion of certain factor spaces, e.g.  domain-wall solutions
considered in earlier paper \cite{IMSjhep}.

Here we considered certain examples of solutions with
exponential dependence of one scale factor and
constant scale factors of internal spaces,
that may be also considered as Freund-Rubin type
solutions generalizing that of \cite{I}.

\begin{center}
{\bf Acknowledgments}
\end{center}

This work was supported in part by the Russian Ministry of
Science and Technology  and Project DFG (436 RUS 113/678/0-1(R)).
V.D.I. is grateful to Prof. S.-W. Kim  for the hospitality
during his stay at Ewha Woman's University, Korea, in March 2004.

\renewcommand{\theequation}{\Alph{subsection}.\arabic{equation}}
\renewcommand{\thesection}{}
\renewcommand{\thesubsection}{\Alph{subsection}}
\setcounter{section}{0}

\section{Appendix}

 \subsection{Equations of motion}

The equations of motion corresponding to  (\ref{2.1}) have the following
form
 \bear{B.4}
 R_{MN}  =   Z_{MN} + \frac{2 V_{pot}}{D - 2} g_{MN},
 \\
 \label{B.5}
 {\btu}[g] \varphi^\alpha -
 \sum_{a \in \Delta} \theta_a  \frac{\lambda^{\alpha}_a}{n_a!}
 e^{2 \lambda_{a}(\varphi)} (F^a)^2
 - 2  \sum_{s \in S_{pot}}   \lambda^{\alpha}_s
 e^{2 \lambda_{s}(\varphi)} \Lambda_s = 0,
 \\
 \label{B.6}
 \nabla_{M_1}[g] (e^{2 \lambda_{a}(\varphi)}
 F^{a, M_1 \ldots M_{n_a}})  =  0,
 \ear
 $a \in \Delta$; $\alpha=1,\ldots,l$.
In (\ref{B.5}) $\lambda^{\alpha}_{a} = h^{\alpha \beta}
 \lambda_{\beta a}$, where $(h^{\alpha \beta})$
is a matrix inverse to $(h_{\alpha \beta})$.
In (\ref{B.4})
 \beq{B.7a}
 Z_{MN}= Z_{MN}[\varphi] +
 \sum_{a \in \Delta} \theta_a e^{2 \lambda_{a}(\varphi)} Z_{MN}[F^a,g],
 \eeq
where
 \bear{B.7}
 Z_{MN}[\varphi] =
 h_{\alpha\beta} \p_{M} \varphi^{\alpha} \p_{N} \varphi^{\beta},
 \\  \label{B.8}
 Z_{MN}[F^a,g] = \frac{1}{n_{a}!}  \left[ \frac{n_a -1}{2 -D}
 g_{MN} (F^{a})^{2}
 + n_{a}  F^{a}_{M M_2 \ldots M_{n_a}} F_{N}^{a, M_2 \ldots M_{n_a}}
 \right].
 \ear

\subsection{Derivation of cosmological solutions}

 Here  we present the derivation of the main solution from the Section 2
 using the general scheme from  \cite{IMJ,IMtop}.

 We consider a metric
 \beq{A.11g}
   g= w e^{2{\gamma}(u)} du \otimes du +
  \sum_{i= 1}^{n} e^{2\phi^i(u)} \hat{g}^i ,
 \eeq
defined on the manifold (\ref{1.2}),
where $g^i$ is a Ricci-flat metric on $M_{i}$
 $i= 1, \dots, n$.

For fields of forms a "composite electro-magnetic"
Ansatz (\ref{1.5}) is adopted with
 \bear{A.28n}
 {\cal F}^s = d\Phi^s \wedge \tau(I_s), \\ \label{A.29n}
  {\cal F}^s = e^{-2\lambda_a(\varphi)}
  *\left(d\Phi^{(a,m,J)}\wedge\tau(I_s)\right),
 \ear
 for $s \in S_e$ and $s \in S_m$, respectively. Here
  $*=*[g]$ is the Hodge operator on $(M,g)$ and forms $\tau(I)$
are defined in (\ref{1.13}).

The potentials in (\ref{A.28n}), (\ref{A.29n})
and scalar fields are functions of $u$, i.e.
 $\Phi^s=\Phi^s(u)$ and  $\varphi^\alpha=\varphi^\alpha(u)$.

Fixing the time gauge to be harmonic one
 \beq{har}
  \gamma = \gamma_0(\phi)
  \equiv \sum_{i= 1}^n d_i\phi^i,
 \eeq
 and  integrating the field equations and Bianchi identities for form
 fields

 \bear{A.max}
  \frac d{du}\left(\exp(-2U^s)\dot\Phi^s\right)=0
     \Longleftrightarrow
      \dot\Phi^s=Q_s \exp(2U^s),
  \ear
 we are led to  Lagrange equations for the Lagrangian
  \beq{A.31n}
   L=\frac12\bar G_{AB}\dot x^A\dot x^B-V,
  \eeq
 with the zero-energy constraint
 \beq{A.33n}
  E=\frac12\bar G_{AB}\dot x^A\dot x^B + V=0.
 \eeq

 Here $(x^A)=(\phi^i,\varphi^\alpha)$,
  \beq{A.32n}
   V= - w \sum_{s \in S_{pot}} \Lambda_s 
     \exp[2\gamma_0 + 2 \lambda_s (\varphi) ]
   + \frac12 \sum_{s\in S_{br}} \eps_s Q_s^2 \exp[2U^s(x)]
  \eeq
is potential and the matrix $(\bar G_{AB})$ is defined in (\ref{2.1}). The
first term in the right hand side of (\ref{A.32n})
corresponds to the scalar potental  and the second one is of a  brane origin.
 Here
  \beq{A.2.3}
   U_A^s x^A =
   \sum_{i \in I_s} d_i x^i - \chi_s \lambda_{s \alpha} x^{\alpha},
   \eeq
  $s \in S_{br}$. The first  term  in the right hand side of
 (\ref{A.32n}) may be rewritten in the "brane form" if  we define
  \beq{A.2.4}
   - w  \Lambda_s = \frac12  \eps_s Q_s^2, \qquad \chi_s = - 1
   \eeq
 for $s \in S_{pot}$.
 Thus, the potential (\ref{A.32n}) reads
  \beq{A.33}
     V=  \frac12 \sum_{s \in S } \eps_s Q_s^2 \exp[2U^s(x)],
  \eeq
where  $S \equiv S_{br} \sqcup S_{pot}$.

For orthogonal co-vectors (\ref{2.7a})
the solutions to Lagrange equations corresponding to  (\ref{A.31n})
read \cite{GIM,IMJ}
 \beq{A.34n}
    x^A(u)=-
    \sum_{s\in S}\frac{U^{sA}}{(U^s,U^s)}\ln |f_s(u-u_s)| + c^A u +
    \bar{c}^A,
  \eeq
where $u_0$, $u_s$ are constants, functions $f_s$ are defined
in (\ref{1.4.5})-(\ref{1.4.9}), $s\in S$, and vectors
 $c^A$,  $\bar{c}^A$ obey the linear constraints
 (\ref{2.8}). The zero-energy restriction (\ref{A.33n})
 is equivalent to the quadratic constraint (\ref{2.10}).

The relations for the metric and scalar fields follow
from the formulas under consideration
and the following useful relations \cite{IMC}
  \beq{A.8n}
    U^{si}= G^{ij} U_j^s = \delta_{iI_s}-\frac{d(I_s)}{D-2}, \quad
    U^{s\alpha}= - \chi_s \lambda_{a_s}^\alpha.
 \eeq

The relations for form fields in Section 2 follow just from (\ref{A.max}).

 \end{document}